\documentstyle[12pt,aps,epsfig]{revtex} 
\begin{document}

\title{The d'-dibaryon in a colored cluster model}  
\author{A.\ J.\ Buchmann, Georg Wagner and Amand Faessler}
\address{Institut f\"{u}r Theoretische Physik, Universit\"at T\"ubingen,
         Auf der Morgenstelle 14, D-72076 T\"ubingen, Germany }

\maketitle

\begin{abstract} 
We calculate the mass and structure of a J$^P$=0$^-$, T=0 six-quark system
using a colored diquark-tetraquark 
cluster wave function and a nonrelativistic quark model Hamiltonian. 
The calculated mass is some 350 MeV above the empirical value
if the same confinement strength as in the nucleon is used.
If the effective two-body confinement strength is weaker in a compound 
six-quark system than in a single baryon, as expected from a simple harmonic 
oscillator model, one obtains $M_{\rm{d'}}$=2092 MeV close to experiment.
\end{abstract}

\section*{Introduction}

In pionic double charge exchange reactions on nuclei at 50 MeV and forward 
angles, there is considerable experimental evidence for a narrow resonance in
the $\pi$NN-system with quantum numbers J$^P$=0$^-$, T=0. 
This resonance has been named $d'$-dibaryon. 
Experimentally, it has a mass of $M_{\rm{d'}}$=2065 MeV, and a free hadronic 
decay width of approximately $\Gamma_{\rm{d'}} \approx$ 0.5 MeV \cite{wag97}.
               
In the present work, we investigate the mass and hadronic structure of a 
J$^P$=0$^-$, T=0 six-quark system in a colored diquark-tetraquark cluster 
model using the Resonating Group Method (RGM) \cite{buc97}. 
This method determines the orbital configuration of the six-quark system 
dynamically, i.\ e.\ according to a given model Hamiltonian. 
Our microscopic approach allows to test the underlying assumption of the 
bag-string model \cite{mul80}, that the $d'$ is a stretched 
diquark-tetraquark system. 
The bag-string model prediction for the $d'$ mass $M_{\rm{d'}}\simeq 2100$ MeV
employs a single, non-antisymmetrized $q^2-q^4$ dumbbell configuration.
The present potential model description of the $d'$ improves 
previous bag-string model calculations in the following respects:
(i) the center of mass energy is exactly removed,
(ii) the Pauli principle for the whole six-quark system is respected,
(iii) by virtue of the antisymmetrizer, the $q^3-q^3$, $q^2-q^4$, and 
$q^1-q^5$ partition into colored clusters, as well as the  $q^6$ compound 
state are automatically included. 
The model accurately reproduces the mass of the deuteron, which is the 
only established dibaryon. 

A major purpose of this work is to study the effect of quark exchange 
interactions (Pauli principle) between the colored clusters on the mass 
and wave function of the $d'$. 
By comparing the RGM solutions with previous quark shell model results 
\cite{wag95} employing a six-quark ``bag'' basis, we obtain additional
information on the amount of clusterization in the system. 
We employ different confinement parametrizations
and study how our results depend on the model of confinement. 
The central question is whether the present model supports
a J$^P$=0$^-$, T=0 state with a mass compatible with experiment.

\section*{Model description} 

The spontaneous breaking of chiral symmetry of low-energy QCD by the physical 
vacuum is responsible for the constituent quark mass generation, as well as 
for the appearence of pseudoscalar and scalar collective excitations of the 
vacuum ($\pi$ and $\sigma$ fields), that couple to the constituent quarks.
The nonrelativistic quark model Hamiltonian for $n$-quarks with equal masses 
$m_q=313$ MeV=$m_N$/3 (in SU$_F$(2)) contains therefore besides the residual 
one-gluon-exchange interaction, modelling asymptotic freedom at short 
distances, and besides the long-range effective two-body confinement potential,
regularized one-pion- and one-sigma-exchange between constituent quarks.
Several two-body confinement potentials, that differ in their radial 
dependence, have been considered \cite{buc97}. 
As usual, the few parameters of the Hamiltonian are fitted to
the nucleon and $\Delta$ ground state masses.

\begin{figure}[h]
\centerline{\epsfig{file=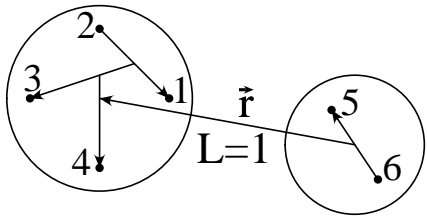, width=5.5cm}}
\caption{The colored diquark-tetraquark cluster model for the $d'$.} 
\label{fig1}
\end{figure}

The six-quark wave function 
$\vert\Psi_{\rm{d'}}\rangle ={\cal{A}} ( \chi_{L=1}({\vec{r}}) \otimes
                               \vert D\rangle\otimes\vert T\rangle )$
sketched in figure \ref{fig1}  
is expanded in the cluster basis into the internal wave functions 
of the tetraquark $\vert T\rangle$ and diquark $\vert D\rangle$ clusters, 
respectively, and the relative wave function $\chi_{L=1}({\vec{r}})$  
between the two clusters, projected onto angular momentum L=1. 
The antisymmetrizer ${\cal{A}}$ of Eq.\ (\ref{antisym}), neglected in previous 
calculations \cite{mul80}, contains the quark permutation operators
$P_{ij}^{XSTC}$ in orbital (X), spin (S), isospin (T) and color-space (C)
\begin{equation}
  {\cal A} = 1 - 8 P_{46}^{XSTC} + 6 P_{35}^{XSTC} P_{46}^{XSTC} \; ,
\label{antisym}
\end{equation}
which ensures that the Pauli principle is respected for the whole six-quark 
system, and allows for a continuous transition from the compound $q^6$ 
six-quark state to the $q^2-q^4$ clusterized state and back.

\section*{Results and Discussion}   

\vspace{0.5cm}
\begin{table}[h]
\caption{We show the diquark $M_D$ and tetraquark masses $M_T$,  
         the dibaryon mass $M_{\rm{d'}}$,  
         the harmonic oscillator parameter $b_6$ minimizing the d' mass, 
         the $d'$ mass neglecting the Pauli principle $M_{\rm{no QEX}}$ 
         and the rms radius for the intercluster coordinate
         $r^{\rm{RGM}}_{\rm{d'}}$ for the two parameter sets 
         discussed in the text.}
\label{table1}
\begin{tabular}{ l  d  d  d  d  d  d }
Set & $M_D$ & $M_T$ & $M_{\rm{d'}}$ & $b_6$ & $M_{\rm{no QEX}}$ &
                                               $r^{\rm{RGM}}_{\rm{d'}}$ \\ 
    & [MeV] & [MeV] & [MeV]         & [fm]  & [MeV]             & [fm]  \\ 
\tableline
I  & 643    & 1456  & 2440          & 0.75  & 2316              & 1.10  \\ 
II & 621    & 1309  & 2092          & 0.95  & 2013              & 1.39  
\end{tabular}
\end{table}

Table \ref{table1} shows our main results for two different
treatments for the effective two-body confinement.
If we assume that the Hamiltonian for baryons and dibaryons is the same,
in particular that there exists a universal effective two-body confinement 
strength for both, baryons and dibaryons (set I), the mass of the 
J$^P$=0$^-$, T=0 state is with $M_{\rm{d'}}$=2440 MeV nearly 400 MeV higher 
than suggested by experiment. 
The use of  different radial functions for the confinement potential 
(e.\ g.\ linear or error-function) leads to some reduction of the predicted 
mass, but $M_{\rm{d'}}$ is still 300-200 MeV higher than the experimental 
resonance position.
A comparison of the results for the $d'$ mass with and without quark exchange 
($M_{\rm{no QEX}}$) show that the quark exchange interactions required by 
the Pauli principle contribute an additional energy of 80-120 MeV.

\begin{figure}[h]
\centerline{\epsfig{file=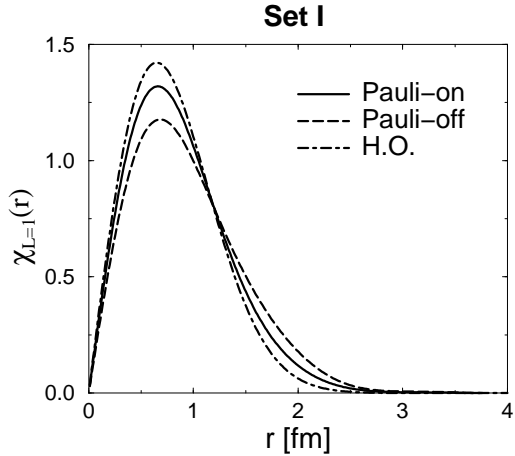, width=7.5cm}}
\caption{The relative RGM wave function between the tetraquark and 
         diquark clusters with (Pauli-on) and without
         (Pauli-off) inclusion of the quark exchange diagrams for set I. 
         The RGM wave functions are compared to a single six-quark shell model 
         state (H.O.).} 
\label{fig2}
\end{figure}

The comparison of the cluster model results \cite{buc97} with previous 
shell model results \cite{wag95} reveals an overall quantitative agreement for 
the $d'$ mass, the orbital structure, and the size parameter $b_6$ of the $d'$. 
We show in figure \ref{fig2}, that the $d'$ wave function does not display a 
pronounced clusterization but resembles rather closely the pure six-quark 
harmonic oscillator (H.O.) wave function given by the dot-dashed curve.
The relative wave function has already died out at distances 
of about 2.5 fm between the clusters. 
The quark exchange diagrams enhance the radial color ``attraction'' between 
the two colored clusters.

The rms radius in the relative cluster coordinate $r^{\rm{RGM}}_{\rm{d'}}$ 
is in general smaller than the sum of the corresponding diquark and 
tetraquark radii. 
In other words, there is considerable overlap between the clusters.
Therefore, the bag-string model assumption of inert colored clusters at the 
ends of a stretched bag \cite{mul80} is not satisfied in the 
present calculation.
The Pauli principle, i.\ e.\ the fact that the quantum numbers of the 
$d'$ are incompatible with two colorless ground state nucleons, together with 
the confinement forces prevent large interquark distances and no distinct
clusterization is observed.  
 
At present, there is no theory of confinement. 
Although we do not know how to calculate the effective confinement strength 
for three- and compound six-quarks systems from first principles, we can gain 
some qualitative insight within the harmonic oscillator model of confinement. 
If we assume a universal confining mean field for any quark in a hadron 
(set II), one derives for the quadratic confinement, that the effective
two-body confinement strength in the six-quark system $a^{(6)}_c$ is 
considerably smaller than in the baryon: 
$a^{(6)}_c \approx a^{(3)}_c/3$.
Besides the dependence of the effective two-body confinement strength on 
the number of quarks and the color representation of the system, the larger
characteristic hadronic size of a six-quark bag as compared to a baryon is 
mostly responsible for the weakening of the six-quark confinement.
This assumption leads to a larger $d'$ and a mass $M_{\rm{d'}}$=2092 MeV 
close to the experimental $d'$ mass.  
Conversely, the empirical $d'$ mass may be interpreted as evidence for a 
weaker effective two-body confinement strength in a compound six-quark system. 

Recently, we have calculated the decay width of the $d'$ \cite{Ito96},
showing that for a  $d'$-mass of $M_{d'} \approx$ 2100 MeV, as predicted
by the weaker confinement hypothesis, the calculated pionic decay width
is compatible with experiment.


\end{document}